# Wall Heat Transfer and Flow Field Configuration of Shock Wave–Turbulent Boundary Layer Interactions on Cryogenically Cooled Wall

Yuma Miki(三木佑真)[1, a)], Leo Ando(安藤嶺央)[1], Azumi Miyazaki(宮崎あずみ)[1],Yasuhiro Egami(江上泰広)[2] and Kiyoshi Kinefuchi(杵渕紀世志)[1]

1 Department of Aerospace Engineering, Nagoya University, Nagoya, Aichi 464-8603, Japan

2 Department of Mechanical Engineering, Aichi Institute of Technology, Aichi 470-0392, Japan

a) Author to whom correspondence should be addressed: miki.yuma.n4@s.mail.nagoya-u.ac.jp

## Abstract

In this study, we experimentally investigated the wall heat transfer and flow field configuration of incident-reflected shock wave–turbulent boundary layer interactions on a cooled wall in supersonic flow. Wind tunnel experiments were conducted at a Mach number of 2.0 and a total temperature of 289 K. To create a cooled-wall state, the wind tunnel wall was cooled to a cryogenic temperature using liquid nitrogen at 77.4 K. In addition to conventional measurements, such as the schlieren visualization method and pressure measurements, cryogenic temperature-sensitive paint was employed to clarify the relationship between the flow field configuration and wall heat flux on a cryogenically cooled wall. The wall surface temperature of the cryogenically cooled wall was 95 K, corresponding to a wall-to-recovery temperature ratio of 0.34. The oil flow image and wall surface temperature distribution indicated a quasi-two-dimensional flow at the center of the wind tunnel. The schlieren images and wall pressure distributions showed that the separation point under the cooled-wall condition shifted downstream compared with that under the uncooled-wall condition. Based on the temperature distribution obtained from the cooled-wall experiments, the wall heat flux at the separation point reduced due to the outward flow from the wall. The peak wall pressure ratio and wall heat flux ratio normalized by their upstream values exhibited trends consistent with previously reported data under the cooled-wall condition. These results suggest that the cryogenic temperature-sensitive paint is a powerful tool for investigating the effects of wall temperature on the shock wave–turbulent boundary layer interactions on cryogenically cooled walls.

# 1. Introduction

Supersonic intakes and isolators in ramjet or scramjet engines decelerate and compress the capture flow through several oblique shock waves[1,2]. These oblique shocks impinge on the boundary layer that develops along the wall. In particular, in the downstream regions of the supersonic intake or at the isolator, the turbulent boundary layer interacts with the oblique shocks, causing a phenomenon known as shock wave–turbulent boundary layer interaction (SWTBLI)[3,4]. SWTBLI often causes undesirable events, such as an increase in the separation bubble that provokes an unstart, an increase in the structural load due to low-frequency unsteadiness, and an increase in the heating load due to strong heat transfer[5,6]. Therefore, it is crucial to gain an accurate understanding of SWTBLIs in order to design high-performance intakes and isolators.

The wall surface temperature, $T_w$, is an important parameter that affects the characteristics of SWTBLIs, such as separation-region size, interaction length, wall pressure distribution, and wall heat flux. The $T_w$ of the actual high-speed vehicles is often lower than the recovery temperature, $T_r$ (adiabatic wall temperature). This means that the wall-to-recovery temperature ratio, $s = T_w/T_r$, is less than one ($s < 1$). Moreover, several future scramjet engines have been proposed, where the walls of the supersonic intakes or isolators at $1 < M < 4$ are actively cooled using cryogenic fuel, thereby realizing high-performance engine cycles[7,8]. Therefore, it is necessary to gain insight into the effects of wall cooling on the SWTBLI from the perspective of wall heat transfer and flow field configuration.

Early experimental studies on SWTBLIs with cooled walls ($s < 1$) were conducted in the 1960s[9–11]. To estimate the wall heat transfer, Markarian[12] reviewed the early studies and concluded that there was a strong correlation between the peak wall heat flux ratio and peak wall pressure ratio: $q_{w,max}/q_{w,u} = (p_{w,max}/p_{w,u})^{0.85}$. Here, $q_w$ and $p_w$ represent the wall heat flux and wall pressure, respectively. The subscripts $max$ and $u$ indicate the peak and reference values upstream of the SWTBLI, respectively. Back et al.[13,14] provided a theoretical explanation for this power law by approximating the energy integral form of the boundary layer. Coleman et al.[15] reported that their model could estimate the $q_w$ distribution from the $p_w$ distribution and mainstream Mach number, $M$. Tang et al.[16] also proposed a more accurate prediction model using $p_w$, $M$, and $s$, compared with the power law and Coleman et al.'s[15] prediction. In summary, these studies[12–16] showed that wall heat flux was strongly related to the wall pressure. However, studies on improving the prediction accuracy have primarily focused on compression ramps in hypersonic flows ($M > 4$)[15,16].

Figure 1 shows the experimental conditions from previous studies, where the mainstream Mach number, $M$, is plotted against the wall-to-recovery temperature ratio, $s$. Both of these parameters are important to improve the wall heat flux prediction and understand the flow field configuration characteristics. However, experiments concerning SWTBLIs on a cooled wall ($s < 1$) in supersonic flows ($1 < M < 4$), corresponding to the flow conditions downstream of the supersonic intake or in the isolator, have only been reported by Spaid et al[17] ($M = 2.9$ and $0.47 \leq s \leq 1.05$) and Back et al.[13,14] ($M = 3.4$ and $s = 0.44$). This is because creating cooled-wall conditions is technically challenging in a supersonic wind tunnel, where the total temperature is the room temperature. Moreover, Spaid et al.[17] and Back et al.[13,14] reported that a decrease in the wall-to-recovery temperature ratio, $s$, induced a reduction in the separation distance, and they did not visualize the differences in the flow field configuration with variations in the wall temperature. To date, experimental studies on cooled walls ($s < 1$) in supersonic flows ($1 < M < 4$) are inadequate to discuss the relationship between wall heat transfer and flow field configuration.

To fill the knowledge gap regarding the effects of wall temperature on SWTBLIs in supersonic flows ($1 < M < 4$), numerical simulation studies have been carried out over the years. Bernardini et al.[18] investigated the effects of wall temperature on the behavior of incident-reflected shock type at $M = 2.28$ and $0.50 \leq s \leq 1.9$ using direct numerical simulation (DNS). They confirmed that the separation region was strongly affected by the wall temperature. In addition, they reported that the peak generation mechanism of the wall heat flux was linked to the turbulence amplification in the interaction region. Many DNS studies on the effects of wall temperature in supersonic flows ($1 < M < 4$) have also been conducted[19–24]. Although there only a few studies that provide experimental data, there is considerable interest in SWTBLIs under the following conditions: $1 < M < 4$ and $s < 1$.

Precise SWTBLI experimental diagnostics on cooled walls ($s < 1$) without disturbing the flow facilitate in accurately understanding the SWTBLI characteristics. Hayashi et al.[25,26] measured the wall heat flux with fast response using multilayered thin film heat transfer gauges. They reported that the fluctuations became strong for an unseparated boundary layer, and the peak appeared near the incident-shock impingement point. In contrast, for a separated boundary layer, significant fluctuations and two peaks occurred throughout the interaction region. Recent experimental studies have been carried out on optical diagnostics. Schülein et al.[27] employed the global interferometry skin friction technique and an infrared radiation (IR) camera, indicating the limit of the Reynolds analogy in the interaction region. Jiao et al.[28] investigated the local enhancement mechanism of the wall heat flux using an IR camera. They found that the spanwise distribution of heat flux was affected by the boundary layer regimes. Lin et al.[29] conducted temperature-sensitive paint (TSP) measurements to investigate the thermal effects generated by incident-reflected SWTBLI. They reported the effects of Görtler vortices on wall heat transfer. In the aforementioned studies[25–29], the researchers performed wall temperature measurements as well as employed the schlieren method. Hence, combining optical measurement methods such as the schlieren method with conventional temperature measurements without disturbing the flow can be used to evaluate the wall heat transfer and flow field configuration, revealing a more detailed picture of the SWTBLI mechanism. However, experimental studies on SWTBLIs using IR or TSP measurements have primarily been conducted on cooled walls in hypersonic flows[25-29].

Meanwhile, experiments on the effects of the wall temperature on SWTBLIs in supersonic flows ($1 < M < 4$) have been conducted under heated-wall conditions ($s > 1$). Jaunet et al.[30] evaluated the SWTBLIs of incident-reflected shock at $M = 2.3$ and $s = 1.0, 1.4, 1.9$ using particle image velocimetry (PIV), schlieren visualization, and time-resolved hot-wire measurements. They discussed the interaction length and low-frequency reflected shock motion of the heated wall. Zhang et al.[31] visualized SWTBLIs at $M = 2.7$ and $s = 1.0, 1.2, 1.4$ using the nanotracer planar laser scattering technique (NPLS) and PIV. They investigated the influence of the wall temperature on the flow field, focusing on detailed vortex structures, boundary layers, and shock wave behavior. Zhong et al.[32] also visualized SWTBLI at $M = 2.95$ and $1.0 \leq s \leq 2.0$ using NPLS and PIV. They not only observed the flow field but also conducted proper orthogonal decomposition and probability density function analyses to compare the mode characteristics of the streamwise velocity fields. In summary, these experimental studies with heated walls focused on the detailed flow characteristics of supersonic flows[30-32] by performing flow field visualization.

In this study, we experimentally investigated the SWTBLIs of incident-reflected shock on a cooled wall in supersonic flows ($1 < M < 4$, $s < 1$), where experimental data are insufficient to discuss the wall heat transfer and flow field configuration, as shown in Fig. 1. Our research group cooled the wall of a supersonic wind tunnel to a cryogenic temperature using liquid nitrogen as the coolant[33]. We confirmed the differences in the flow field configuration of the SWTBLI between cooled- and uncooled-wall conditions. However, the flow-field measurement was limited to schlieren visualization, and no quantitative data on heat transfer characteristics were obtained. This was because heat flux sensors could not be installed in this wind tunnel without disturbing the flow, and the IR cameras could not work because of the low radiation intensity from the cryogenically cooled wall.

The purpose of this study is to evaluate the relationship between the wall heat transfer and flow field configuration of incident-reflected SWTBLIs on the cryogenically cooled wall, by employing cryogenic temperature-sensitive paint (cryoTSP)[34-36] to measure the surface temperature. CryoTSP can optically measure the cryogenic wall temperature distribution with high spatial resolution without disturbing the flow. We conducted wall pressure measurements, the schlieren visualization, oil flow visualization, and cryoTSP measurements to investigate the relationships between the wall temperature, wall heat flux distribution, and flow field inside the SWTBLIs. Furthermore, the wall pressure ratio, $p_{w,max}/p_{w,u}$, and wall heat flux ratio, $q_{w,max}/q_{w,u}$, were compared with the power law and previously reported data.

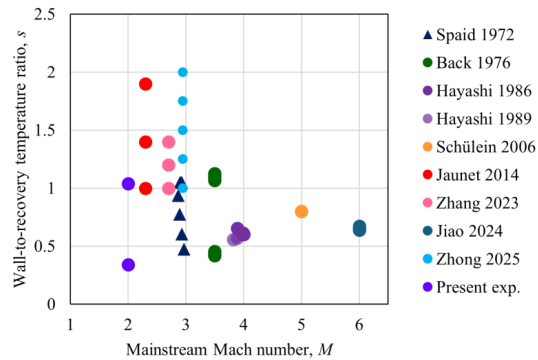

Fig. 1. Previous experimental conditions of SWTBLIs. The triangle markers represent compression ramp flow, whereas the circle markers represent incident-reflected shock flow. The wall-to-recovery temperature ratio, $s = T_w/T_r$ means the wall temperature condition compared to the case of the adiabatic wall. $s < 1$, $s = 1$, $s > 1$ means cooled wall, adiabatic wall, and heated wall, respectively.

## 2. Experimental Setup and Procedure

### 2.1. Supersonic Wind Tunnel

Figure 2 shows the schematic of the atmospheric air breathing-type supersonic wind tunnel. The mainstream Mach number in the test section was 2.0. Atmospheric air was used as the test gas, which was dried using a silica gel dryer at the inlet of the wind tunnel. The total pressure and total temperature in the test section were 101 kPa and 289 K, respectively. The cross-sectional area of the test section was rectangular, with a height of 38 mm and width of 80 mm. The upper wall was flat and made of an aluminum alloy (A6063) with a thickness of 2.0 mm. The upper wall was cooled by pouring liquid nitrogen at 77.4 K into a pool installed on its exterior surface. Polycarbonate was used for the side and bottom walls to prevent cooling.

To observe the SWTBLIs on the cooled and uncooled walls, a 13° wedge with a width of 79 mm was installed on the bottom wall to generate an oblique shock wave. This oblique shock wave interacted with the turbulent boundary layer on the upper wall. There is enough distance from the wind tunnel inlet to the test section to develop fully turbulent boundary layer. Therefore, the SWTBLIs of the incident-reflected shock occurred on both the cooled and uncooled walls with or without liquid nitrogen.

Static pressure holes and K-type thermocouples were installed on the upper wall of the wind tunnel,

as shown in Fig. 2(b) and listed in Table I. Figure 3 shows a cross-sectional view of the test section. The pressure tubes extending from these measurement points were connected to pressure gauges (Panasonic DP-101ZA, estimated accuracy: $\pm 1.15$ kPa) via the liquid nitrogen pool. The K-type thermocouples were fixed to the exterior surface of the upper wall.

To evaluate the flow field, schlieren visualization was conducted in the cooled and uncooled wall experiments with the knife-edge direction set horizontally. Hence, the recorded images exhibited a density gradient in the vertical direction. To record schlieren movies, we used a high-speed camera (Phantom Inc. V1211) at 100 frames per second along with BK-7 windows and a light source (Cavitar Ltd., CAVILUX Smart) with a pulse duration of 40 ns.

To capture the three-dimensional structure of the flow, oil flow measurements were performed only under the uncooled-wall condition. We could not obtain an oil flow image under the cooled-wall condition because the oil had solidified on the cryogenically cooled wall. Instead of the oil flow image, we assessed the three-dimensionality of the cooled-wall condition using cryoTSP measurements.

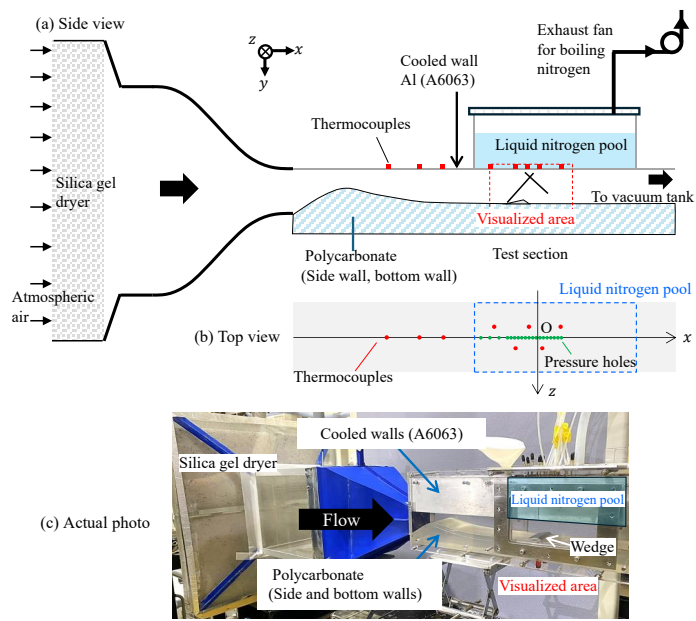

Fig. 2. Schematic of the supersonic wind tunnel. Here, eight red and 20 green points denote the positions of the thermocouple and pressure holes, respectively: (a) side view (b) top view and (c) actual photo. We set the $x$-coordinate origin at the point of impingement of the oblique shock on the upper wall. The $y$-coordinate represents the normal distance from the upper wall. The $z$-coordinate origin was at the center of the wind tunnel span.

Table I. Positions of the thermocouples. Thermocouples $T_{c1}$–$T_{c3}$ were installed upstream of the liquid nitrogen pool, whereas thermocouples $T_{c4}$–$T_{c8}$ were installed in the liquid nitrogen pool.

|  | $T_{c1}$ | $T_{c2}$ | $T_{c3}$ | $T_{c4}$ | $T_{c5}$ | $T_{c6}$ | $T_{c7}$ | $T_{c8}$ |
|---|---|---|---|---|---|---|---|---|
| $x$ (mm) | −188 | −148 | −118 | −58 | −23 | −8 | 7 | 37 |
| $z$ (mm) | 0 | 0 | 0 | −15 | 15 | −15 | 15 | −15 |

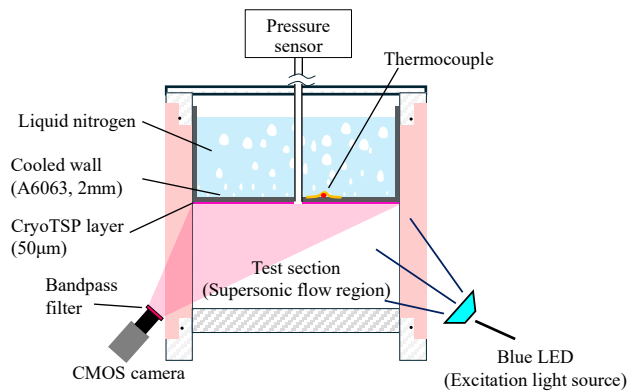

Fig. 3. Cross-sectional view of the test section with pressure sensor, thermocouple, and cryoTSP.

## 2.2. Cryogenic Temperature-Sensitive Paint

The wall surface temperature in the SWTBLI region on the upper wall was measured using cryoTSP. CryoTSP measurement is suitable for measuring the cryogenic wall surface temperature without disturbing the flow when the use of IR cameras would be impractical. For the TSP measurements, the TSP containing a luminophore is applied to the model surface. When illuminated at an appropriate wavelength, this luminophore emits fluorescence whose intensity varies with temperature. Therefore, by comparing the luminous intensities of a reference image and a result image captured by a camera, we could calculate the temperature distribution of the model surface with high resolution[36].

In this work, we applied cryoTSP with a polyurethane binder and Ru(trpy)$_2$ as a temperature-sensitive luminophore to the inner surface of the upper wall[34–36]. The measurement system is illustrated in Fig. 3. A blue light-emitting diode (LED) with a central wavelength of 470 nm was used as the excitation light source for Ru(trpy)$_2$. We used a 12-bit monochrome complementary metal oxide semiconductor (CMOS) camera (Toshiba Teli Corp., BG505LMG) equipped with a $600 \pm 40$ nm bandpass filter to capture the upper wall, recording the luminous intensity distribution at the emission peak wavelength of 605 nm. The CMOS camera was set at a frame rate of 20 Hz and exposure time of 49 ms. We adopted an intensity-based method to calculate the temperature distribution. To reduce random noise, a reference image was obtained by averaging five images acquired immediately before the wind tunnel test, assuming a uniform wall temperature. By comparing the reference image with the images recorded during the wind tunnel test, we could calculate the temperature distribution using a calibration curve that correlated the luminous intensity with temperature. The calibration curve and temperature of the reference image were obtained by installing a thermocouple on the cryoTSP layer during precooling. The temperature measurement accuracy was estimated to be approximately 1.5–3.0 K from the random shot noise of the CMOS camera.

## 2.3. Experimental Procedure

Under the cooled-wall condition, liquid nitrogen was added to precool the upper wall. Figure 4 shows the wall temperature histories measured using the thermocouples throughout the cooling experiment. The horizontal axis represents time $t$, which was measured from the start of the wind tunnel test. One hundred and fifty seconds after the beginning of the precooling operation, the wind tunnel test was initiated when the upper wall temperature in the test section reached a steady state. During the wind tunnel running, the upper wall temperature upstream of the liquid nitrogen pool ($T_{c1} - T_{c3}$) gradually increased, whereas the temperature within the liquid nitrogen pool ($T_{c4} - T_{c8}$) remained steady.

For the uncooled-wall case, during the wind tunnel running, the upper wall temperature remained virtually at room temperature.

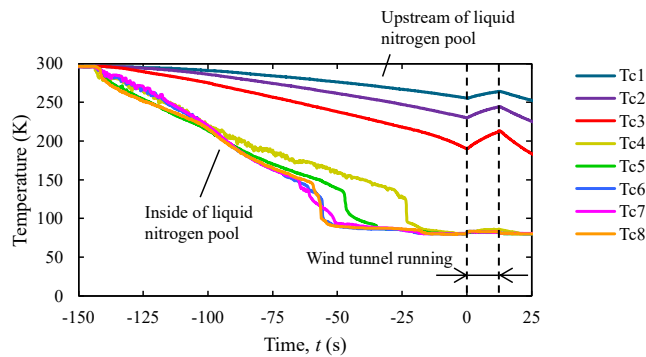

Fig. 4. Temperature histories of the upper wall under the cooled-wall condition. Thermocouples $T_{c1} - T_{c3}$ were installed upstream of the liquid nitrogen pool, whereas thermocouples $T_{c4} - T_{c8}$ were installed within the liquid nitrogen pool.

# 3. Results and Discussion

## 3.1. Flow Establishment Under the Cooled-Wall Condition

Figure 5 shows the wall surface temperature distribution, $T_w$, obtained from the cryoTSP measurements. In Fig. 5, the flow direction is from left to right, and the top and bottom edges correspond to the sidewalls. The dots at the span center ($z = 0$) represent the static pressure holes. In the interaction region ($x > -20$ mm), where the oblique shock impinges on the boundary layer on the cooled wall, the wall surface temperature in each frame differed from the upstream temperature at the center and near the sidewalls. In addition, at the center of the wind tunnel ($-15$ mm $\leq z \leq$ 15 mm), the temperature distribution was almost constant along the spanwise direction.

The temperature histories at four points on the inner upper wall surface obtained from the cryoTSP measurements are plotted in Fig. 6, along with the temperature histories obtained from the thermocouples installed at the bottom surface of the liquid nitrogen pool. The four points of the cryoTSP histories are the spatial-averaged values in 5 mm × 5 mm areas at $(x, z) = (-48$ mm, $\pm$ 10 mm$), (0$ mm, $\pm 10$ mm$)$. The temperature histories obtained from the cryoTSP measurements showed an abrupt jump immediately after the wind tunnel started. After this abrupt jump, the four points' temperatures remained relatively constant ($1.4$ s $\leq t \leq 2.0$ s). Therefore, the cryoTSP measurements indicated that the SWTBLI at the center of the wind tunnel ($-15$ mm $\leq z \leq$ 15 mm) was an approximately steady flow for $1.4$ s $\leq t \leq 2.0$ s.

Figure 7 shows the instantaneous schlieren and digital streak images under the cooled-wall condition. In the schlieren image, the flow direction is from left to right, and the incident shock from the bottom wedge interacted with the turbulent boundary layer on the upper wall. The history of the flow field can be evaluated from the digital streak image, which displays the pixel rows at $x = -48$ mm and $-16$ mm of each schlieren frame in chronological order[37]. Although the gradual temporal change of the flow field was caused by an increase in the upper wall temperature ($T_{c1} - T_{c3}$) upstream of the liquid nitrogen pool (Fig. 4), the incoming boundary layer and separation region remained approximately steady when $1.4$ s $\leq t \leq 2.0$ s, as indicated by Figs. 7(a) and 7(b), respectively. Since the wall temperature and flow field were confirmed to be approximately steady, we regarded $1.4$ s $\leq t \leq 2.0$ s as the test time to evaluate the flow field configuration and wall heat transfer.

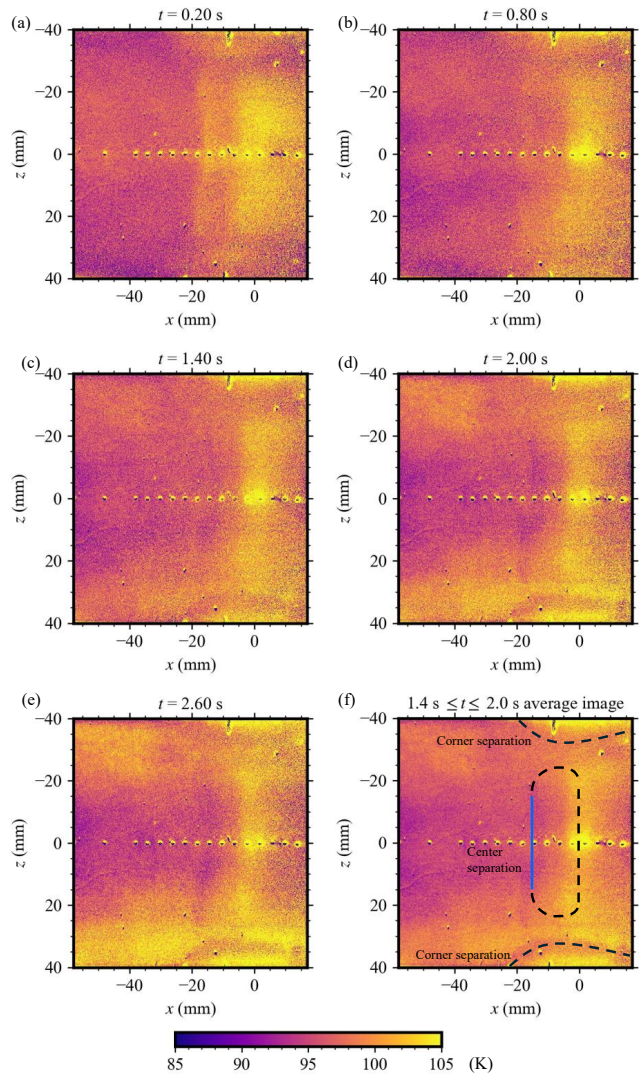

Fig. 5. Temperature distributions obtained from the cryoTSP measurements under the cooled-wall condition: (a)–(e) results of each frame and (f) time-average temperature distribution during the test time ($1.4\ \mathrm{s} \leq t \leq 2.0\ \mathrm{s}$). Solid bule line was estimated from the schlieren image as shown in Fig. 11(a). Dashed lines were approximate position predicted from the temperature distribution.

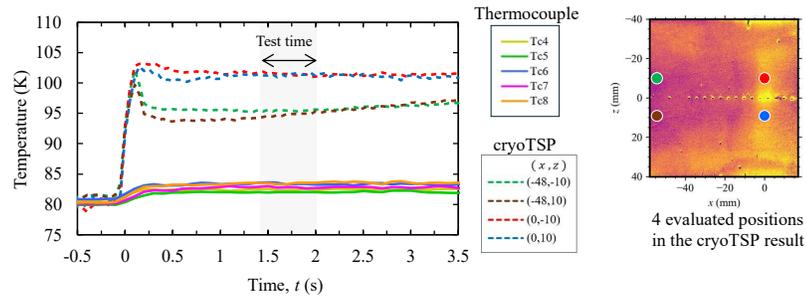

4 evaluated positions
in the cryoTSP result

Fig. 6. Temperature histories obtained from the thermocouple and cyroTSP measurements in the test section during the wind tunnel operation. The solid lines represent the temperature histories obtained from the thermocouples. The dashed lines represent the temperature histories obtained from the cryoTSP measurements at four points. The temperature during the test time ($1.4 \text{ s} \leq t \leq 2.0 \text{ s}$) was approximately steady.

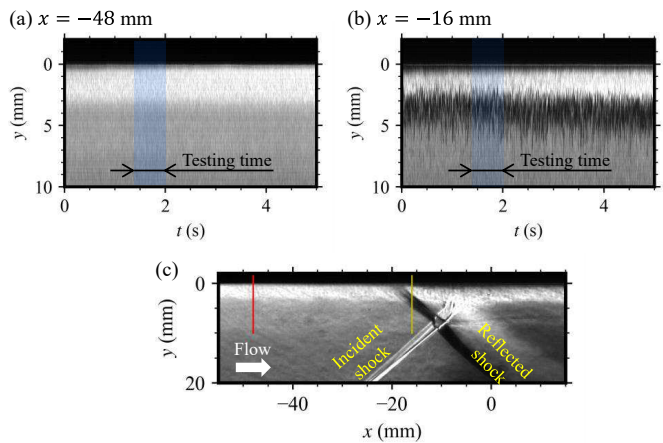

Fig. 7. Digital streak images showing the time histories of the flow field on the cooled wall measured using the schlieren method: (a) incoming boundary layer history at $x = -48$ mm, (b) separation shock history at $x = -16$ mm, and (c) instantaneous schlieren image. The red and yellow lines indicate the locations of the digital streak images at $x = -48$ mm and $-16$ mm. A digital streak image is generally generated by arranging the pixel rows of each schlieren image in chronological order.

## 3.2. Wall-To-Recovery Temperature Ratio

The wall surface temperature range, $T_w$, in our experiment was 95–105 K during the test time. The recovery temperature is expressed as

$$T_r = T_e \left( 1 + r \frac{\gamma - 1}{2} M_e^2 \right), \qquad (1)$$

where $T_e$ and $M_e$ are the static temperature and Mach number of the mainstream, respectively. The specific heat ratio of air, $\gamma$, was set at 1.4. The recovery factor, $r$, was set at 0.896. The $r$ of the turbulent boundary layer of the airflow was calculated from $r = \sqrt[3]{Pr}$, where Prandtl number $Pr$ was 0.72. Therefore, the recovery temperature, $T_r$, was 276 K because $T_e = 165$ K and $M_e = 2.0$.

The wall-to-recovery temperature ratio, $s$, under the cooled-wall condition was 0.34 at $T_w = 95$ K obtained from the cryoTSP measurements. Therefore, we confirmed that this condition resulted in the lowest Mach number and wall-to-recovery temperature ratio ever reported in experimental studies on SWTBLIs, as shown in Fig. 1.

For the uncooled-wall condition, the wall surface temperature, $T_w$, was approximately equal to the room temperature (289 K) due to the low level of thermal convection. Hence, the wall-to-recovery temperature ratio was estimated as $s = 1.04$.

## 3.3. Flow field configuration

When evaluating the flow field configuration of the SWTBLIs from the schlieren image, it is important to consider three-dimensionality. Figure 8 shows the schlieren and oil flow visualization images under the uncooled-wall condition. The separation line was observed not only in the center of the wind tunnel, but also near the sidewalls, as reported by Xian and Babinsky[38,39]. They found that the compression and expansion waves from the corner separations can distort the center separation line along the spanwise direction. However, the center separation line at $-15\text{ mm} \leq z \leq 15\text{ mm}$ in Fig. 8 was virtually linear, indicating that the corner effects were minor. Moreover, the position of the center separation line, $x_s$, was estimated from the schlieren image by assuming that the luminous intensity at the separation line was lower than that in the incoming boundary layer. Hence, the position of the center separation line was found to be $x_s = -25$ mm, as estimated from the schlieren and oil flow visualization images.

Although we could not obtain an oil flow visualization image for the cooled-wall condition, the cryoTSP measurement results suggest that the corner effects were minor as a substitute for the oil flow image. This is supported by the nearly constant temperature distribution along the spanwise direction at $-15\text{ mm} \leq z \leq 15$ mm, as described in Sec. 3.1 and shown in Fig. 5 (f). Thus, we regarded the flow fields at the center region ($-15\text{ mm} \leq z \leq 15$ mm) under both cooled- and uncooled-wall conditions as a quasi-two-dimensional flow with the SWTBLIs of the incident-reflected shock.

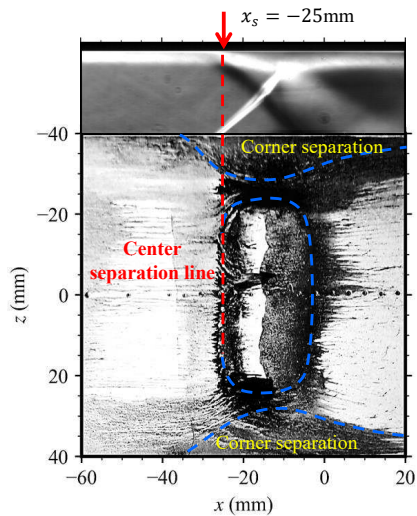

Fig. 8. Oil flow and schlieren images under the uncooled-wall condition. The top and bottom edges corresponded to the sidewalls. The center separation line, marked in red, was located at $x_s = -25$ mm.

Figure 9 shows the time-averaged schlieren images and wall pressure distributions, $p_w$, under both uncooled- and cooled-wall conditions during the test time. Figure 10(a) shows a schematic of the detailed flow field configuration along with the wall pressure distribution. The incident shocks from the bottom wedge interacted with the turbulent boundary layer, causing SWTBLIs. In both conditions, the pressure rise onset position corresponded to the separation shock foot in the schlieren images. The schlieren images and wall pressure distributions showed good agreement with each other, indicating that both measurements accurately captured the flow field configuration. The incident shock angle and wall pressure distributions at $x \leq -30$ mm were identical for the uncooled- and cooled-wall conditions, indicating that the mainstream flow upstream of the interaction region remained unchanged. This is because heat transfer from the upper wall affects only the boundary layer.

Next, we explain the differences in the flow field configuration between the two conditions. The positions of the pressure rise onset and foot of the separation shock $x_0$ shifted downstream as the wall temperature decreased. The interaction length, $L_{int}$, under the cooled-wall condition reduced compared with that under the uncooled-wall condition, where $L_{int}$ is given by $x_{imp} - x_0 = -x_0$, as illustrated in Fig. 10(b). Here, $x_{imp}$ is the shock impingement position and the $x$ coordinate origin in this study. The boundary layer thickness, $\delta$, at $x = -48$ mm was estimated to be 4.0–4.5 mm on the uncooled wall and 3.8–4.2 mm on the cooled wall, based on the schlieren images.

On the other hand, interestingly, the pressure distributions downstream of $x = -6.0$ mm were almost identical. We inferred that the pressure wave in the mainstream was in the same state.

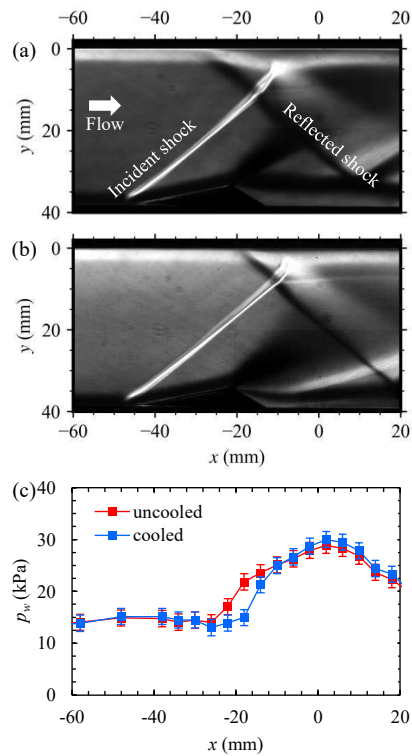

Fig. 9. Time-averaged schlieren images and wall pressure distributions during the test time: (a) uncooled-wall condition and (b) cooled-wall condition. (c) Wall pressure distributions of the cooled- and uncooled-wall conditions. The error bars indicate the standard errors.

Babinsky and Harvey[4] noted that the dependence on wall temperature exceeded the scope of the free interaction theory[40]. The shift mechanism of the separation shock and pressure rise onset position should be considered by focusing on the interaction length, $L_{int}$, as illustrated in Fig. 10. The following interpretations of $L_{int}$ are important: the distance that the information from the incident shock propagates upstream through the subsonic channel of the boundary layer, and the level of the adverse pressure gradient, $-\Delta p_t/L_{int}$, experienced by the boundary layer, where $\Delta p_t$ is defined as $p_{w,max} - p_{w,u}$, as shown in Fig. 10. Based on Sabnis and Babinsky's[41] explanation, a decrease in the height of the subsonic region within the boundary layer results in a decrease in the interaction length, $L_{int}$. In contrast, an increase in the adverse pressure gradient, $-\Delta p_t/L_{int}$, induces an increase in the height of the subsonic region. Hence, it is thought that these two factors, the height of the subsonic region and resistance of the incoming boundary layer to the adverse pressure gradient, jointly influence the interaction length, $L_{int}$.

The wall cooling on the boundary layer reduces the height of the subsonic region and increases the resistance to the adverse pressure gradient. This is because the static temperature and viscosity decrease, whereas the density increases. The reduction in viscosity slightly makes velocity profile fuller and an increased velocity gradient near the wall. Consequently, the skin friction coefficient and momentum flux in the vicinity of the wall increase, and the thickness of the subsonic region decreases. Based on this mechanism, we inferred that the decrease in the wall-surface temperature, $T_w$ (i.e., the decrease in the wall-to-recovery temperature ratio, $s = T_w/T_r$), caused the reduction in the interaction length, $L_{int}$, in this experiment.

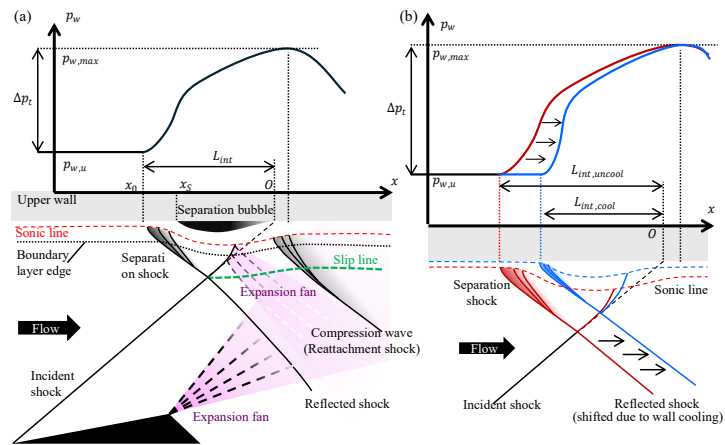

Fig. 10. (a) Schematic of the common flow field configuration with wall pressure distribution. The pressure jump $\Delta p_t$ and the incident shock were maintained constant for both uncooled- and cooled-wall conditions. (b) Differences in wall pressure distributions and shock positions between the uncooled- and cooled- wall conditions. Red and blue lines indicate the uncooled- and cooled-wall cases, respectively.

## 3.4. Relationship Between the Wall Heat Transfer and Flow Field Configuration

In this section, we discuss the relationship between the wall heat transfer and flow field configuration under the cooled-wall condition. Figures 11(a), (b), and (c) show the schlieren image, nondimensional wall heat flux distribution $q_w/q_{w,u}$, and nondimensional wall pressure distribution $p_w/p_{w,u}$, respectively. The upstream reference values, $q_{w,u}$ and $p_{w,u}$, were set to be the values at $x = -48$ mm.

The wall heat flux, $q_w$, during the test time was estimated by assuming the wall heat transfer as a one-dimensional steady thermal conduction problem, as illustrated in Fig. 12. Hence, $q_w$ can be expressed as

$$q_w = R(T_w - T_c). \tag{2}$$

Here, $T_w$ is the wall surface temperature of the upper wall inside the wind tunnel obtained from the cryoTSP measurements. $T_c$ is the wall surface temperature at the bottom of the liquid nitrogen pool, which was measured by the thermocouples. $R$ represents the thermal resistance of the upper cooled wall. The assumption of steady thermal conduction is reasonable because $T_w$ and $T_c$ remained constant during the test time, as shown in Fig. 6.

The streamwise distribution of $T_w$ was obtained by spanwise averaging the temperature measured using cryoTSP over the ranges $-15$ mm $\leq z \leq -2.5$ mm and 2.5 mm $\leq z \leq 15$ mm. As explained in Sec. 3. 2, the region $-15$ mm $\leq z \leq 15$ mm was considered to be a quasi-two-dimensional flow. The distribution of the pool's bottom-wall temperature $T_c$ was assumed to be uniform at 82.8 K, because the variations among $T_{c4} - T_{c8}$ upstream of and within the interaction region were negligible, as shown in Fig. 6. The spatial distribution of $R$ was also uniform. Hence, the nondimensionalized wall heat flux, $q_w/q_{w,u}$, can be expressed as

$$\frac{q_w}{q_{w,u}} = \frac{R(T_w - T_c)}{R(T_{w,u} - T_c)} = \frac{T_w - T_c}{T_{w,u} - T_c}. \tag{3}$$

Therefore, the wall heat flux ratio, $q_w/q_{w,u}$, can be calculated from $T_w$ and $T_c$, which were measured using the cryoTSP and thermocouples, respectively.

The wall pressure and wall heat flux distributions were similar to each other, except at the separation point ($x_s = -16$ mm), as shown in Figs. 11(b) and 11(c). At $x_s = -16$ mm, $q_w/q_{w,u}$ decreased and $p_w/p_{w,u}$ increased. Interestingly, whether the wall heat flux at the separation point decreases or not differed in the literature. In experimental studies, Hayashi[25] and Jiao[28] reported results similar to our experiment. In addition, most numerical simulation studies[16,18,20,22,42–44] reported a decrease in wall heat flux at the separation point. In contrast, experimental studies of Coleman[15] and Schülein[27] did not report a reduction in wall heat flux at the separation point. Understanding the mechanism of the wall heat flux distribution, $q_w$, is crucial not only for revealing the fluid physics, but also for improving the fluid structure–thermal analysis[6,45].

Tang *et al.*[16,44] explained the cause of the wall heat flux reduction by analyzing hypersonic SWTBLIs over compression ramps ($M = 5, s = 0.5$). Based on their explanation, the decrease in the wall heat flux at the separation point is mainly due to the flow deflection away from the wall at the separation point, as shown in Fig. 13. Consequently, the aerodynamic heat is mainly transported to the mainstream, leading to a decrease in the wall heat flux. We believe that the dominant mechanism of our results qualitatively agrees with Tang *et al.*'s[16,44] explanation. This is because the reduction in the wall heat flux began at the separation shock foot ($x \cong -20$ mm in Figs. 11(a) and 11(b)), where the wall-normal outward flow component started to emerge. In addition, the location of minimum $q_w$ was close to the separation point at $x_s = -16$ on the schlieren image.

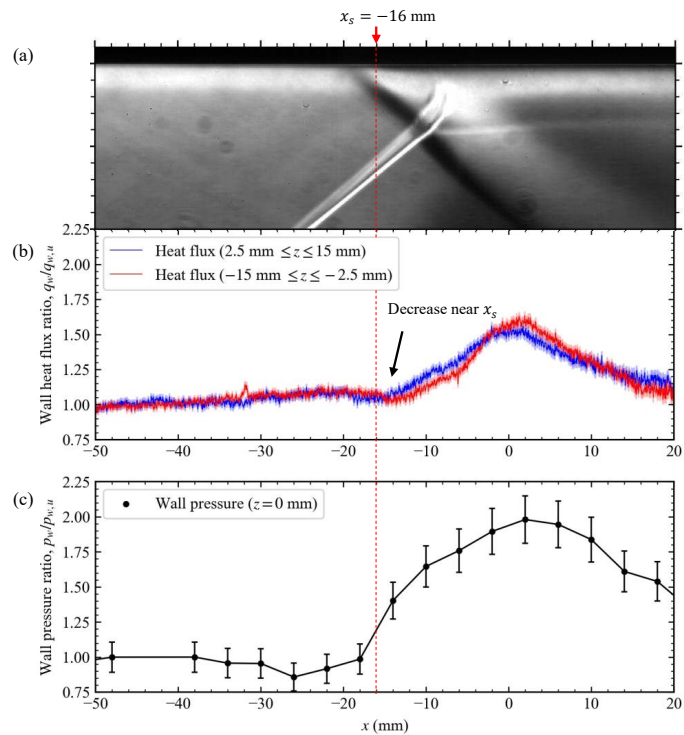

Fig. 11. Experimental results of the cooled-wall condition: (a) time-averaged schlieren image, (b) nondimensional wall pressure, $p_w/p_{w,u}$, at $z = 0$ mm, and (c) nondimensional wall heat flux, $q_w/q_{w,u}$, averaged over $-15$ mm $\leq z \leq -2.5$ mm and $2.5$ mm $\leq z \leq 15$ mm. The line thickness in (b) and the error bars in (c) represent the standard errors. $x_s = -16$ mm was the separation point identified from the schlieren image, where the wall heat flux reduced.

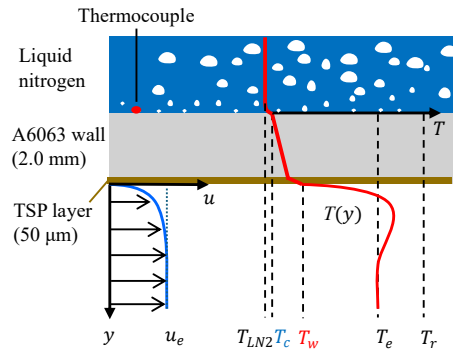

Fig. 12. Wall heat transfer of the cooled wall. The liquid nitrogen temperature, $T_{LN2}$, was 77.4 K. $T_c$ and $T_w$ denote the temperatures measured by the thermocouples and cryoTSP, respectively.

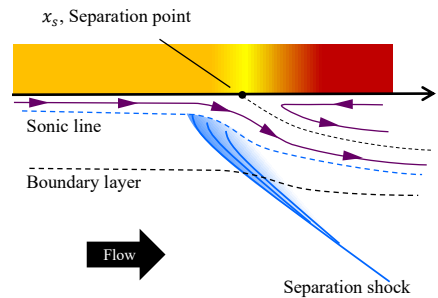

Fig. 13. Schematic of the streamlines at the separation point $x_s$. The flow deflection away from the wall leads to a reduction in wall heat transfer.

The wall heat flux ratio and wall pressure ratio reached their peak values at the same position, as shown in Figs. 11(b) and 11(c), respectively. Figure 14 shows the relationship between the peak wall heat flux ratio and peak wall pressure ratio in incident-reflected SWTBLIs, including our results and those of previously reported numerical and experimental studies. The local enhancement in the wall heat flux was correlated with the peak wall pressure. Even though the power law $q_{w,max}/q_{w,u} = (p_{w,max}/p_{w,u})^{0.85}$ is the accepted prediction model, as described in Section 1, our results better fit the following power law $q_{w,max}/q_{w,u} = (p_{w,max}/p_{w,u})^{0.75}$. The exponent value of 0.75 was the value estimated by Volpiani's simulation for the cooled-wall case[42]. In addition, Zuo[46] reported that the exponent varied within a range of 0.75–0.95 based on the researcher's Reynolds-averaged Navier–Stokes simulations. Recently, some studies suggested that the exponent for a cooled wall was lower than that of a heated wall[16,42]. This trend was indeed observed in Fig. 14. The reason is inferred that the Reynolds shear stress and turbulent heat flux became stronger as the wall temperature increases when the pressure jump is constant, as reported by Bernardini *et al.*[18] As a conclusion, our present result agreed with the previously reported trend that the exponent of the power law is smaller as the wall-to-recovery temperature ratio, $s$, decrease.

To the best of the authors' knowledge, there is no prediction model that is more accurate than the power law reported for incident–reflected SWTBLIs. To further improve the prediction accuracy of the peak heat flux ratio, $q_{w,max}/q_{w,u}$, not only the wall-to-recovery temperature ratio, $s$, but also the mainstream Mach number and the upstream and downstream boundary layers of the incident–reflected shock should also be considered, as in Tang *et al.*'s [16] prediction model for compression ramp flows. More accurate models can be developed to predict the peak wall heat flux through further experimental, numerical, and theoretical investigations under a wider range of flow conditions.

Additionally, to understand the wall heat transfer in more detail, it is necessary to characterize not only the time-averaged wall heat flux and its relationship with the flow field, but also the temporal fluctuations of wall heat flux and the associated flow structures. In particular, understanding the coupling between low-frequency shock oscillations[47], coherent structures such as Görtler vortices[29], and wall heat flux is essential. The present experimental system provides a test duration of 0.6 s, which is sufficient long time assess the kilohertz-order low-frequency dynamics of the SWTBLI. High-speed schlieren methods can be implemented by increasing the camera frame rate. The time response of TSP is generally limited by thermal diffusion within the coating layer. By employing fast-response TSP techniques, including higher frame rates and a thinner TSP layer, temperature histories on the sub-millisecond timescale could be obtained[36,48]. Therefore, the influence of wall temperature on low-frequency shock motion can be quantitatively evaluated by conducting high-speed schlieren methods and fast-response cryoTSP.

In conclusion, this experimental approach can provide valuable data for advancing the understanding of wall heat transfer and unsteady flow dynamics, complementing previous numerical and theoretical studies over a wider range of flow conditions.

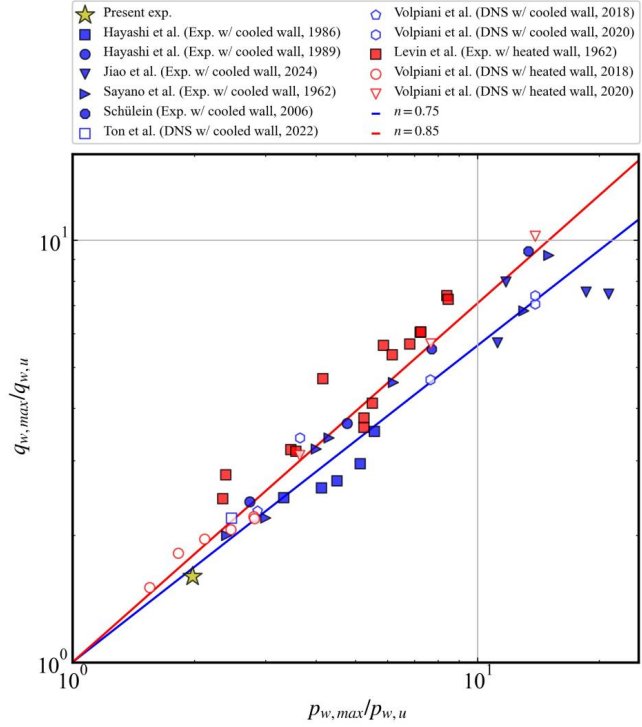

Fig. 14. Peak wall heat flux ratio and peak wall pressure ratio in incident-reflected SWTBLIs. Each data marker indicates the following: (1) filled data markers: values from experiments, (2) hollow data markers: values from numerical simulations, (3) red data markers: heated wall $(s > 1)$, and blue data markers: cooled wall $(s < 1)$. The red and blue solid lines represent the power law $q_{w,max}/q_{w,u} = (p_{w,max}/p_{w,u})^n$, where $n = 0.85$ and $0.75$, respectively.

## 4. Conclusion

In this study, we conducted experiments with cooled and uncooled walls to investigate the relationship between the wall heat transfer and flow field configuration of incident-reflected SWTBLIs at a mainstream Mach number of 2.0 and total temperature of 289 K. An oblique shock was generated using a 13° wedge installed on the bottom wall of the wind tunnel to induce SWTBLIs on the upper wall. For the cooled-wall experiment, the upper wall was cooled cryogenically using liquid nitrogen.

Wall pressure measurements and schlieren visualization were conducted under both cooled- and uncooled-wall conditions. Oil flow visualization and cryoTSP measurements were performed only for the uncooled- and cooled-wall conditions, respectively. For the cooled-wall case, only the test time, which was approximately regarded as the steady period, was analyzed.

CryoTSP successfully captured detailed surface temperature distributions on the cryogenic wall, yielding wall temperatures of approximately 95–105 K, compared with 289 K for the uncooled wall. The corresponding wall-to-recovery temperature ratios were 0.34 and 1.04, respectively, indicating the lowest combination of Mach number and wall-to-recovery temperature ratio reported to date for experimental SWTBLI studies.

The oil flow visualization conducted under the uncooled-wall condition revealed boundary layer separation at the center of the wind tunnel and near the sidewalls. The center separation line agreed with that estimated from the schlieren image. Although oil flow visualization was not carried out for the cooled wall, the cryoTSP measurements indicated negligible spanwise temperature variations near the center of the wind tunnel. Based on these observations, the flow fields in the central region were regarded as quasi-two-dimensional under both cooled- and uncooled-wall conditions.

The schlieren images and wall pressure distributions were consistent under both cooled- and uncooled-wall conditions. For the cooled-wall condition, the separation point shifted downstream and the interaction length reduced compared with that for the uncooled-wall condition. This was attributed to the thinner incoming boundary layer, which increased the resistance to the adverse pressure gradient. In contrast, the peak wall pressure remained relatively invariant because the total pressure jump imposed by the mainstream shock was similar in both cases.

The wall heat flux on the cooled wall was estimated by assuming steady one-dimensional heat conduction. The wall heat flux distribution generally followed the wall pressure trend, except near the separation point, where the heat flux decreased. This observation qualitatively agreed with Tang *et al.*'s[16,44] explanation that the outward flow at the separation point reduces wall heat transfer. In our experiments, the peak wall pressure ratio and peak wall heat flux ratio, normalized by their upstream values, conformed with the power law $q_{w,max}/q_{w,u} = (p_{w,max}/p_{w,u})^{0.75}$. The exponent of 0.75 was lower than Back *et al.*'s[13,14] theoretical value of 0.85. This was likely due to the suppression of the Reynolds stress and turbulent heat flux at lower wall-to-recovery temperature ratios, as suggested by Bernardini *et al.*'s[18] DNS studies.

Our results demonstrated that cryoTSP is a powerful tool for investigating SWTBLIs on cryogenically cooled walls. Further experiments under various conditions will provide valuable data for clarifying the effects of the wall temperature on the SWTBLI heat transfer and flow field configuration.

**Acknowledgments**

This work was supported by JSPS KAKENHI Grant Numbers 25K22131 and by JST SPRING, Grant Number JPMJSP2125. Yuma Miki, would like to take this opportunity to thank the "THERS Make New Standards Program for the Next Generation Researchers." The authors would also like to appreciate Professor Akihiro Sasoh for his valuable advice.

**Author declaration**

*Conflict of interest*

The authors have no conflicts to disclose.

*Ethics approval*

Not applicable

**Data availability**

The data that support the findings of this study are available from the corresponding author upon reasonable request.